\documentclass[preprint,12pt]{elsarticle}
\usepackage{amssymb}
\usepackage{graphicx}

\journal{Intermetallics}

\begin{document}

\author[INT]{M. J. Winiarski}
\address[INT]{Institute of Low Temperature and Structure Research, Polish Academy of Sciences, Ok\'olna 2, 50-422 Wroc\l aw, Poland, EU}

\title{Electronic structure of non-centrosymmetric superconductors Re$_{24}$(Nb;Ti)$_5$ by \emph{ab initio} calculations.}

\begin{abstract}
Electronic structures of superconducting Re$_{24}$Nb$_5$ and Re$_{24}$Ti$_5$ have been calculated employing the full-potential local-orbital method within the density functional theory. The investigations were focused on the influence of the antisymmetric spin-orbit coupling on band structures and Fermi surfaces of these non-centrosymmetric systems. The predicted here density of states at the Fermi level for Re$_{24}$Ti$_5$ is abnormally low with respect to that deduced from previous heat capacity measurements. This discrepancy suggests an intermediate coupled superconducting state in Re$_{24}$Ti$_5$. The differences between electronic properties of both compounds could explain more robust superconductivity in the Nb-based material.
\end{abstract}


\maketitle

\section{Introduction}
The superconducting properties in Re$_{24}$(Nb;Ti)$_5$ intermetallics, adopting the $\alpha$-Mn type structure \cite{Trzebiatowski}, have been studied since 1960s \cite{Matthias1,Matthias2,Steadman}. The recent renewal of an interest in this family of compounds \cite{Lue_Nb,Karki,Lue_Ti} is connected with a search of unusual properties of a superconducting state in non-centrosymmetric systems. The lack of spatial inversion symmetry leads to the antisymmetric spin-orbit coupling (ASOC) that lifts the Krammers degeneracy and may induce a mixture of spin singlet and triplet states in the superconducting phase \cite{Edelstein,Gorkov,Frigeri,Samokhin}.

Despite the fact that in Re-based compounds the strength of ASOC is expected to be relatively strong, the superconducting state in  Re$_{24}$Nb$_5$ (T$_{SC}$ = 8.7 K \cite{Lue_Nb}), Re$_{23.8}$Nb$_{5.2}$ (T$_{SC}$ = 8.8 K \cite{Karki}), and Re$_{24}$Ti$_5$ (T$_{SC}$ = 5.8 K \cite{Lue_Ti}) exhibits rather a BCS-like behaviour. This effect is believed to be related to the quality of polycrystalline samples used in particular studies or still too weak influence of the ASOC on bands forming the Fermi surface (FS) in this family of compounds \cite{Lue_Ti}. 

In this work, we present the results of fully relativistic calculations of electronic structure for Re$_{24}$Nb$_5$ and Re$_{24}$Ti$_5$. They allow for a general discussion of an ASOC strength in the $\alpha$-Mn type Re-based intermetallics as well as an analysis of differences in FS topology and the density of states (DOS) at the Fermi level (E$_F$), the Sommerfeld coefficient ($\gamma$), and the estimated electron-phonon enhancement factor ($\lambda_{ep}$) in these non-centrosymmetric superconductors. 

\section{Computational details}
Band structure calculations have been carried out with the full-potential local-orbital (FPLO) code \cite{FPLO}. The Perdew-Wang parametrisation (PW92 \cite{JP}) of the local density approximation (LDA) was employed in the scalar and fully relativistic modes. Experimental values of lattice parameters of the $\alpha$-Mn type (space group I$\bar{4}$3{\it m}, no. 217) unit cell (u.c.), $a$ = 0.96076 nm \cite{Lue_Nb} and 0.9581 nm \cite{Lue_Ti} were used for Re$_{24}$Nb$_5$ and Re$_{24}$Ti$_5$, respectively. The atomic positions were fully optimised. Valence-basis sets were automatically selected by the internal procedure of FPLO-9. Total energy values of considered systems were converged with accuracy to ~1 meV for the 8x8x8 {\bf k}-point mesh of the Brillouin zone (BZ), corresponding to 87 {\bf k}-points in the irreducible part of the BZ.

\section{Results and discussion}
The calculated positions of Re and Nb/Ti atoms in the $\alpha$-Mn type u.c. of Re$_{24}$Nb$_5$ and Re$_{24}$Ti$_5$ are presented in Table \ref{table1}. Compared to the available experimental data for Re$_{3}$Nb \cite{Steadman}, the results seem to be reasonable. Further experimental studies on single crystals are required to confirm these LDA-based predictions.

\begin{figure}
\includegraphics[width=8cm]{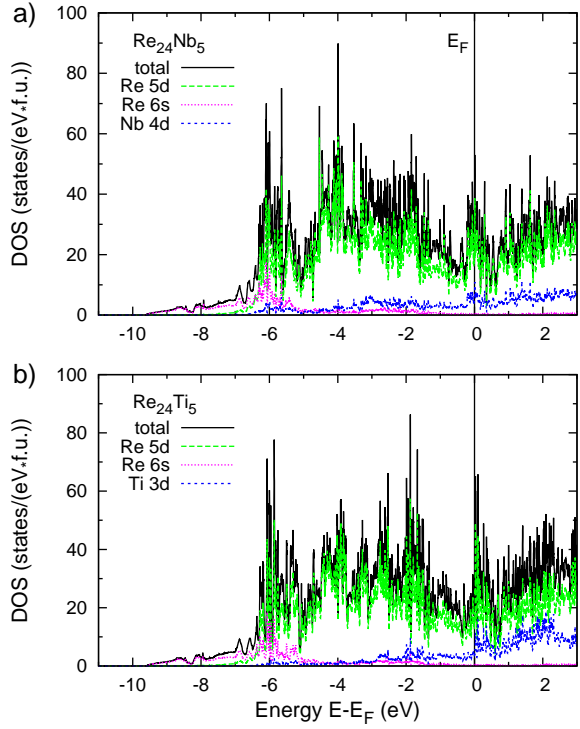}
\caption{The total DOSs of Re$_{24}$Nb$_5$ a) and Re$_{24}$Ti$_5$ b) with partial DOS contributions originating form some states of Re, Nb, and Ti atoms, calculated in the fully relativistic approach.}
\label{Fig1}
\end{figure}

For both studied here compounds, the overall shape of DOS in the vicinity of the Fermi level, presented in Fig. \ref{Fig1}, is completely dominated by the Re 5d states. The contributions of Nb/Ti 4d/3d states become significant at E$_F$ and for higher energies. Meanwhile, the minor contributions of the Re 6s electrons are located about 6 eV below E$_F$.  It is worth noting that the Fermi levels in Re$_{24}$(Nb;Ti)$_5$ systems are located almost at the local maximum of DOS. This feature may enhance the superconducting properties in these intermetallics. The calculated value of N(E$_F$) = 30.00 states/eV/f.u. for Re$_{24}$Nb$_5$ is higher than N(E$_F$) = 21.96 states/eV/f.u. for Re$_{24}$Ti$_5$, mainly due to the bigger contribution of the Nb 4d states in the former compound than in the case of the Ti 3d electrons in the latter one. Since the formula unit of these intermetallics is relatively large, such low values of N(E$_F$) suggest a weakly metallic character of both superconductors.

The calculated here value of the Sommerfeld coefficient, $\gamma_e$ = 70.7 mJ/mol/K$^2$ for Re$_{24}$Nb$_5$, is in a satisfying agreement with the experimental  value, $\gamma$ = 106.8 mJ/mol/K$^2$ for Re$_{23.8}$Nb$_{5.2}$ \cite{Karki}. Since electronic correlations in studied here compounds are rather negligible \cite{Lue_Ti}, the value of electron-phonon enhancement factor, $\lambda_{ep}$ $\approx$ 0.5, derived from the relation: $\gamma = \gamma_e ( 1 + \lambda_{ep})$, should be a reasonable estimation. It suggests a weakly coupled superconducting state in Re$_{24}$Nb$_5$, in accord with the former experimental studies \cite{Lue_Nb}.

Unexpectedly, the calculated $\gamma_e$ = 53.4 mJ/mol/K$^2$ for Re$_{24}$Ti$_5$ differs substantially from experimental $\gamma$ = 111.8 mJ/mol/K$^2$, reported by Lue et al. \cite{Lue_Ti}. In this case, the calculated value of electron-phonon enhancement factor, $\lambda_{ep}$ $\approx$ 1.2, suggests an intermediate coupling and leads to a discrepancy between presented here results and experimental $\lambda_{ep}$ $\approx$ 0.6 from ref. \cite{Lue_Ti}. This issue requires some further discussion.
The relatively low value of calculated N(E$_F$) could be caused by an artificial position of E$_F$ since, as seen in Fig. \ref{Fig1}, a slight shift of E$_F$ to higher energies may lead to a significant increase of N(E$_F$). It is a well known fact, that the LDA approach usually yields an underestimation of u.c. volumes. Thus, the used here experimental values of the lattice parameter {\it a} could influence the results as compressive strain. However, this possibility has been excluded by the calculations for the decreased lattice parameter {\it a} by 1 \%, leading to a significantly lower value of N(E$_F$) in Re$_{24}$Ti$_5$. Furthermore, the results were also confirmed by calculations for a denser {\bf k}-point grid to eliminate some numerical inaccuracy.
On the other hand, the discrepancy could also be related to the particular polycrystalline samples of Re$_{24}$Ti$_5$ as well as the fitting procedure for specific heat data. Thus, further experimental studies on single crystals of this compound might confirm the previous results.

\begin{figure}
\includegraphics[scale=0.60,angle=-90]{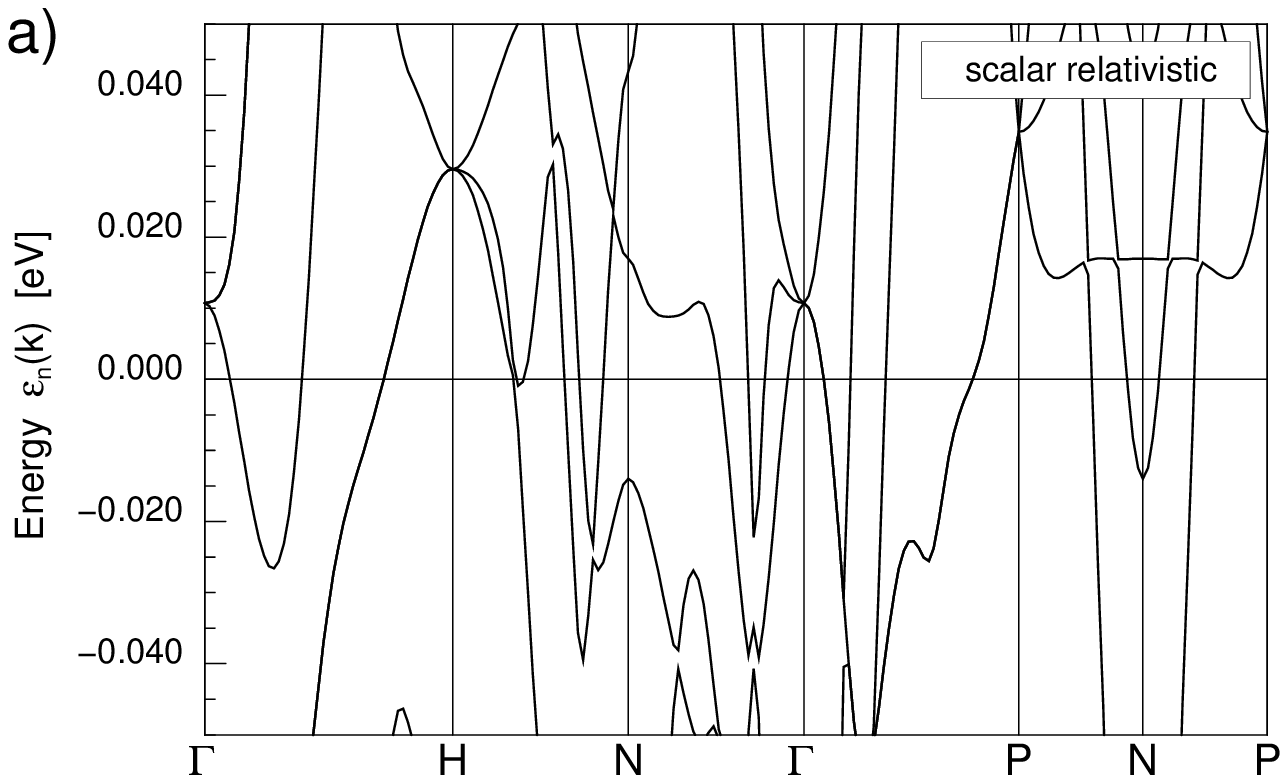}
\includegraphics[scale=0.60,angle=-90]{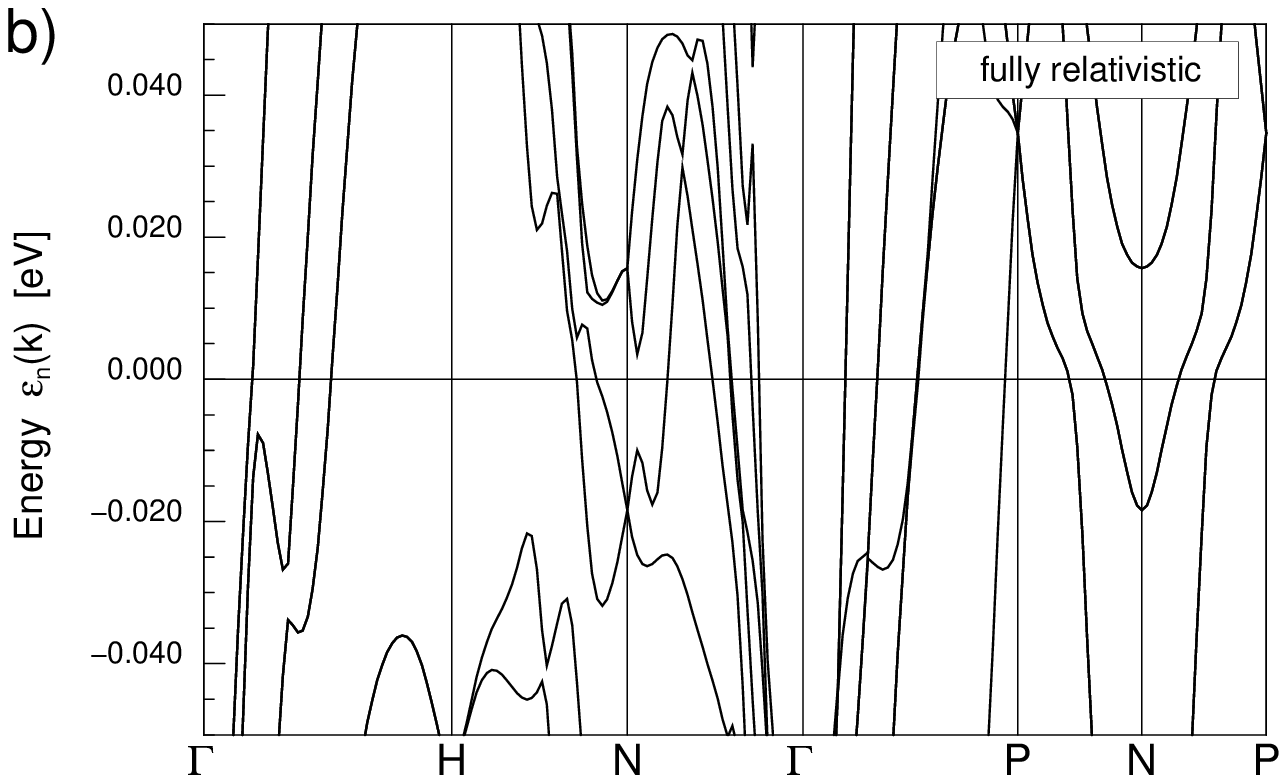}
\caption{Calculated scalar a) and fully b) relativistic band structures of Re$_{24}$Nb$_5$ in the vicinity of Fermi level.}
\label{Fig2}
\end{figure}

\begin{figure}
\includegraphics[scale=0.60,angle=-90]{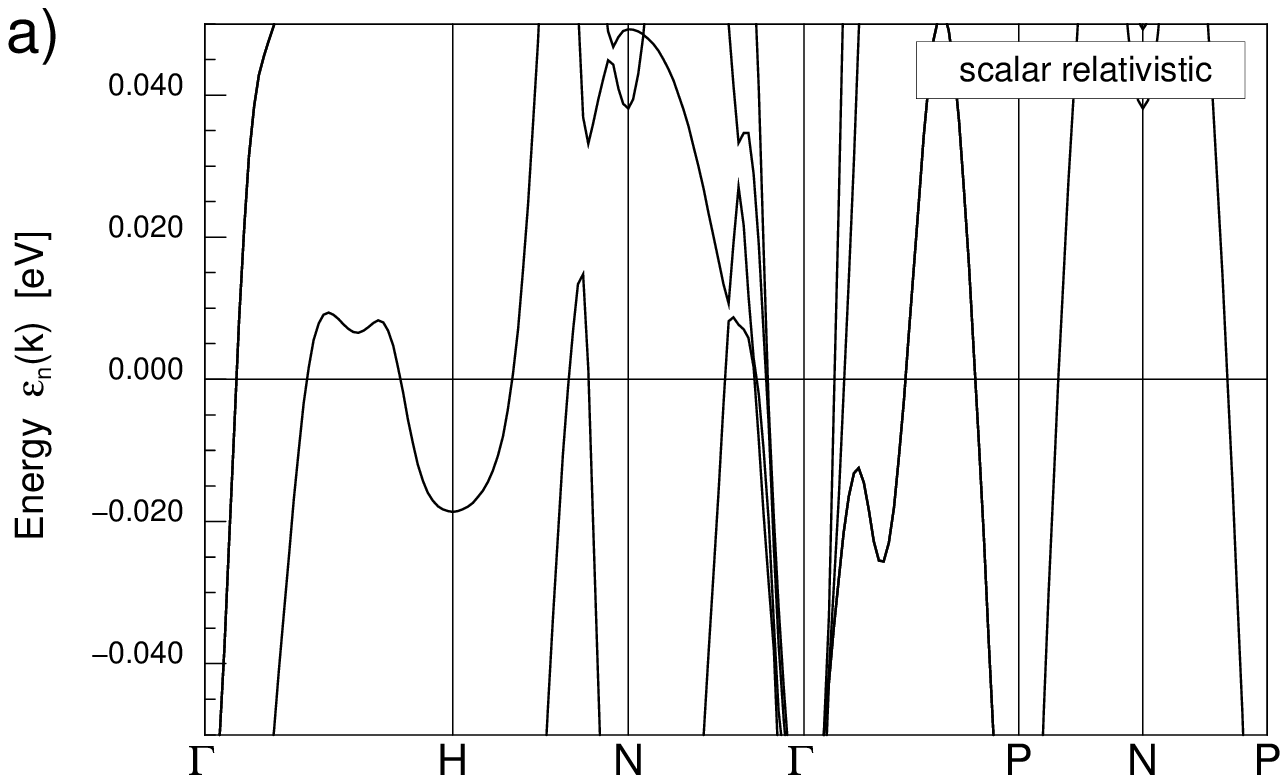}
\includegraphics[scale=0.60,angle=-90]{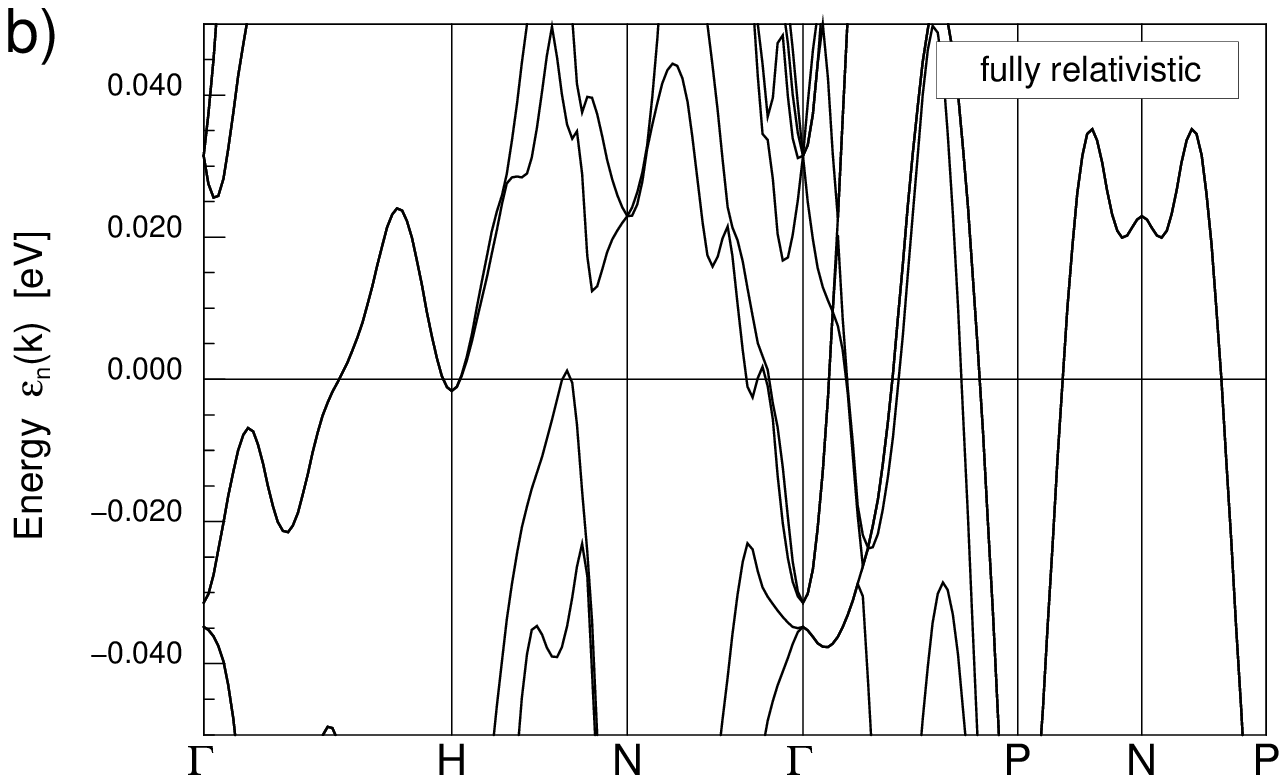}
\caption{The same as in Fig. \ref{Fig2} but for Re$_{24}$Ti$_5$.}
\label{Fig3}
\end{figure}

The influence of ASOC on the electronic band structure of Re$_{24}$Nb$_5$ is presented in Fig. \ref{Fig2}. The positions of scalar relativistic bands in the vicinity of E$_F$ are strongly changed after employing the fully relativistic approach, leading to a different topology of FS. Only the bands in the $N$-$P$ line of BZ remained almost unmodified. Some analogous but weaker effects could be also seen in bandplots for Re$_{24}$Ti$_5$, depicted in Fig. \ref{Fig3}. However, in the Ti-based compound a relatively lower number of bands crosses E$_F$, while the topology of its FS is similar in both scalar and fully relativistic approaches.

\begin{figure}
\includegraphics[width=8cm]{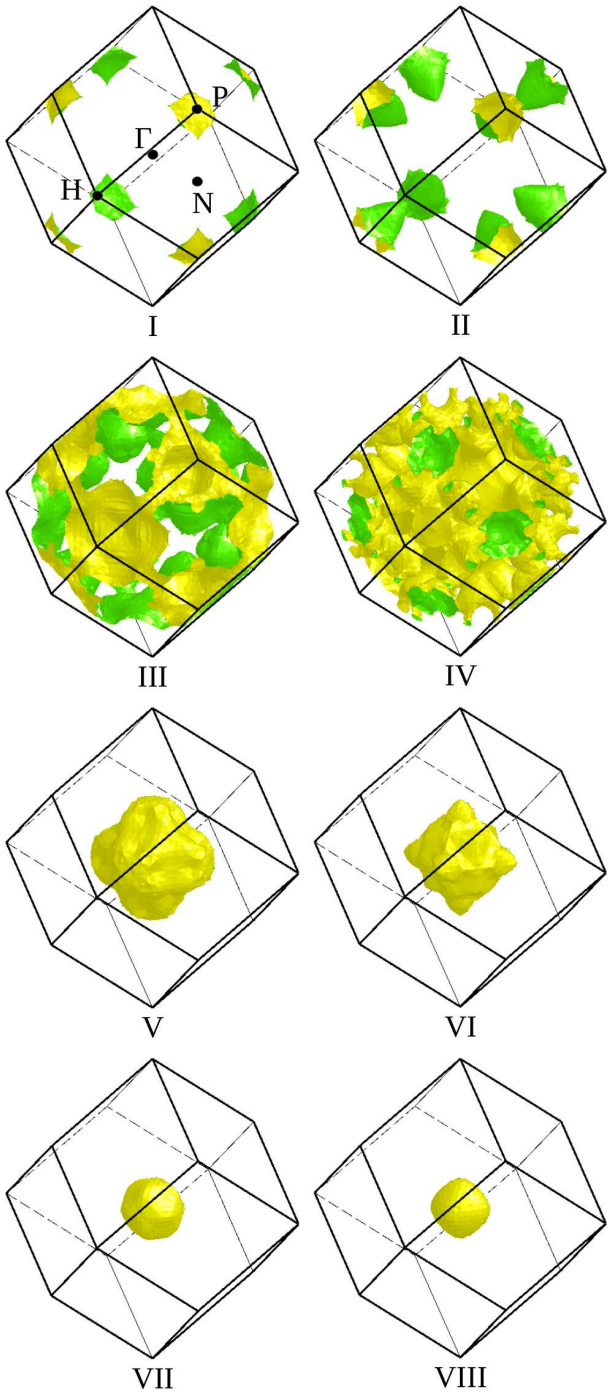}
\caption{Holelike I-III and electronlike IV-VIII Fermi surface sheets of Re$_{24}$Nb$_5$, calculated in a fully relativistic mode.}
\label{Fig4}
\end{figure}

\begin{figure}
\includegraphics[width=8cm]{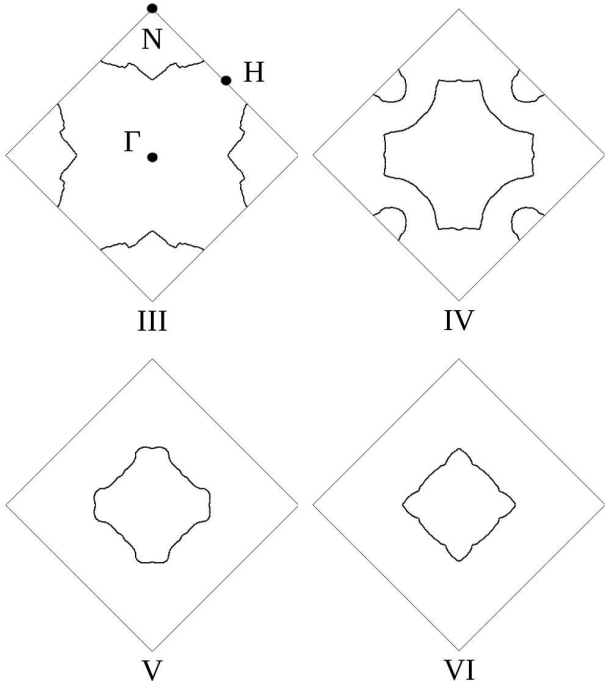}
\caption{Sections of the Fermi surface in the $\Gamma$-$H$-$N$ plane for sheets III-VI of Re$_{24}$Nb$_5$, presented in Fig \ref{Fig4}.}
\label{Fig5}
\end{figure}

The calculated FS sheets of Re$_{24}$Nb$_5$, presented in Fig. \ref{Fig4}, exhibiting holelike (I-III) and electronlike (IV-VIII) character, form pairs of similar topology. Substantial differences have been revealed only in pair III-IV of sheets, as depicted in Fig. \ref{Fig5}. The character of sheet III is holelike while sheet IV is rather electronlike. In turn, the driven by ASOC modifications of bands preserve the highly symmetric shape of the FS in Re$_{24}$Nb$_5$. The almost spherical symmetry of FS in this compound may also explain a good correspondence between the properties of its superconducting state and the BCS theory.

\begin{figure}
\includegraphics[width=16cm]{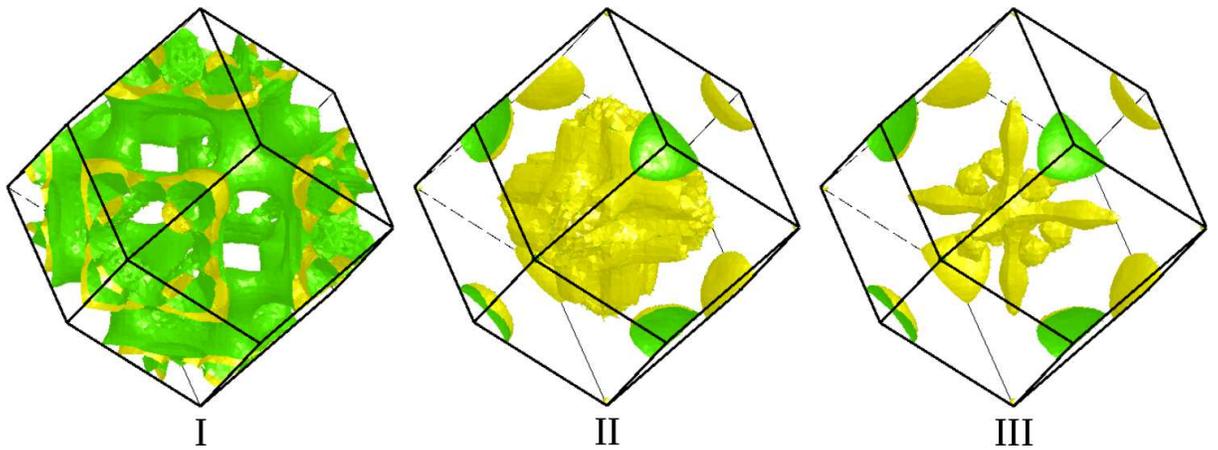}
\caption{Holelike I and electronlike II-III Fermi surface sheets of Re$_{24}$Ti$_5$, calculated in the fully relativistic approach.}
\label{Fig6}
\end{figure}

The FS of Re$_{24}$Ti$_5$ is depicted in Fig. \ref{Fig6}. The holelike sheet I is single while electronlike sheets II and III are very similar. Despite that the topology of FS in the Ti-based compound is generaly different from that of the Nb-based system, one can notice some similarities between these sheets and sheets  V-VI of FS in Re$_{24}$Nb$_5$. It is worth noting that the free atoms of Ti and Nb are not isovalent, the former possesses only 2 electrons on the d-shell while the latter possesses 4 electrons of this type. Since the d-electrons dominate N(E$_F$) in binary Re-Nb/Ti intermetallics, presented in this work differences in electronic structure of Re$_{24}$Nb$_5$ and Re$_{24}$Ti$_5$ have been expected.

\section{Conclusions}
Electronic structures of non-centrosymmetric superconductors Re$_{24}$Nb$_5$ and Re$_{24}$Ti$_5$ have been studied from first principles. The DOS at the Fermi level in both compounds is dominated by the Re 5d electrons. The calculated value of the Sommerfeld coefficient for the Ti-based compound disagrees with the available experimental data, suggesting an intermediate coupled superconductivity in this material. This issue requires a confirmation by further measurements for single crystals.

The modifications of band structures, related to the antisymmetric spin-orbit coupling, are clearly visible for both studied superconductors. However, the Fermi surface in Re$_{24}$Nb$_5$ preserves a highly symmetric character of the majority of four pairs of ASOC double-split sheets. In turn, the FS in Re$_{24}$Ti$_5$, consisting of only 3 sheets, is different from that in the former system.

Presented here findings encourage to further investigations on the relations between electronic structure and superconductivity in the $\alpha$-Mn type intermetallic compounds.

\section*{Acknowledgements}
The author acknowledges M. Samsel-Czeka\l a for valuable discussions. The calculations were performed in Wroclaw Centre for Networking and Supercomputing (Project No. 158).

\begin{table}
\caption{Calculated atomic positions in unit cells of Re$_{24}$Nb$_5$ and Re$_{24}$Ti$_5$ in comparison with experimental data for Re$_3$Nb (taken from ref. \cite{Steadman}).}
\label{table1}
\begin{tabular}{llllll}
atom & site & x/a & y/b & z/c \\ \hline
Re$_{24}$Nb$_5$: & & & & \\
Nb & 2a & 0 & 0 & 0 \\
Nb & 8c & 0.3178 & 0.3178 & 0.3178 \\
Re & 24g & 0.3590 & 0.3590 & 0.0419 \\
Re & 24g & 0.0916 & 0.0916 & 0.2854 \\ \hline
Re$_{24}$Ti$_5$: & & & & \\
Ti & 2a & 0 & 0 & 0 \\
Ti & 8c & 0.3159 & 0.3159 & 0.3159 \\
Re & 24g & 0.3594 & 0.3594 & 0.0419 \\
Re & 24g & 0.0913 & 0.0913 & 0.2840 \\ \hline
Re$_{3}$Nb \cite{Steadman}: & & & & \\
Nb & 2a & 0 & 0 & 0 \\
Nb & 8c & 0.314 & 0.314 & 0.314 \\
Re & 24g & 0.360 & 0.360 & 0.040 \\
Re & 24g & 0.096 & 0.096 & 0.277 \\ \hline
\end{tabular}
\end{table}

\end{document}